\documentclass[aps,prl,twocolumn,groupedaddress]{revtex4-1}

\usepackage{bm}
\usepackage{amsmath}
\usepackage{color,soul}
\usepackage{graphicx}
\usepackage{booktabs}
\newcommand{\ra}[1]{\renewcommand{\arraystretch}{#1}}

\usepackage{dcolumn}
\newcolumntype{d}[1]{D{.}{.}{#1}}

\usepackage[normalem]{ulem}

\newcommand{\vcc}[1]{{\boldsymbol{#1}}}

\newcommand{\abs}[1]{\lvert #1 \rvert}

\renewcommand{\eqref}[1]{(\ref{#1})}

\newcommand{\komma}{\, \mathrm{,}}

\renewcommand{\eqref}[1]{(\ref{#1})}

\newcommand{\eg}{e.g.,\@ }
\newcommand{\ie}{i.e.\@ }
\newcommand{\cf}{cf.\@ }
\newcommand{\etal}{\textit{et al.\@ }}

\begin{document}


\title{Influence of complex disorder on skew-scattering Hall effects in $L1_0$-ordered FePt alloy}


\author{Bernd Zimmermann}
\author{Nguyen H. Long}
\author{Phivos Mavropoulos}
\author{Stefan Bl\"{u}gel}
\author{Yuriy Mokrousov}
\affiliation{Peter Gr\"{u}nberg Institut and Institute for Advanced Simulation, Forschungszentrum J\"{u}lich and JARA, 52425 J\"{u}lich, Germany}

\date{\today}

\begin{abstract}
We show by first-principles calculations that the skew-scattering anomalous Hall and spin-Hall angles of L$1_0$-ordered FePt drastically depend on different types of disorder. A different sign of the AHE is obtained when slightly deviating from the stoichiometric ratio towards the Fe-rich side as compared to the Pt-rich side. For stoichiometric samples, short-range ordering of defects has a profound effect on the Hall angles and can change them by a factor of $2$ as compared to the case of uncorrelated disorder. This might explain the vast range of anomalous Hall angles measured in experiments, which undergo different preparation procedures and thus might differ in their crystallographic quality.
\end{abstract}


\maketitle


Future information technology will heavily rely on spin-orbit effects, which enable the all-electric control of magnetization and spin-degrees of freedom. Spin currents already play a vital role in state-of-the-art technology, for example in spin-transfer torque magnetic access memories (STT-MRAM), and will become ever more important in emergent magnetic technologies. Bright prospects of relativistic spin currents are associated in particular with their key importance for the phenomena of spin-orbit torque \cite{Liu+Buhrmann:Torque-switching-SHE-Ta}, current-induced domain wall \cite{ryu2013chiral}
and skyrmion motion \cite{Hoffmann:blowing_magnetic_skyrmions}, and ultrafast magnetic applications \cite{kampfrath2013terahertz}.

At the heart of spin-orbit transport effects lie the anomalous and spin Hall effects (AHE and SHE) \cite{Spin-Hall-review:Sinova}, because they allow for an efficient conversion from a longitudinal charge current (that is, aligned parallel to an applied electric field) into a transverse charge  and  spin current, respectively. For these microscopically spin-orbit coupling (SOC) originated phenomena there is already a relatively established knowledge of their underlying mechanisms, which partly root in topological properties, thus fundamentally relating the
AHE and SHE to the physics of~e.g.~skyrmions \cite{nagaosa2013topological}, orbital magnetism \cite{thonhauser2011theory} and topological metals \cite{PhysRevB.86.115133}. Conventionally, three relatively distinct contributions to the AHE and SHE are discussed: the so-called intrinsic Berry phase contribution stemming from the electronic structure of a pristine crystal, and two contributions which arise due to disorder, namely, the side-jump and skew-scattering \cite{RevModPhys.82.1539}. Among the three, it is the skew scattering which dominates the Hall effects in the limit of small disorder. The reason is the linear scaling of the skew-scattering driven transverse conductivity $\sigma_{xy}$ with the diagonal conductivity $\sigma_{xx}$ for vanishing scattering. The corresponding scaling constants, the so-called anomalous or spin Hall angles, AHA or SHA, are respectively defined as $\alpha^{\rm AHE} = \sigma^{\rm c}_{xy}/\sigma^{\rm c}_{xx}$ and $\alpha^{\rm SHE} = \sigma^{\rm s}_{xy}/\sigma^{\rm c}_{xx}$, where superscripts ``c'' and ``s'' refer to the charge and spin conductivity tensors, respectively.

\begin{figure}[t!]
	\includegraphics[width=0.49\textwidth]{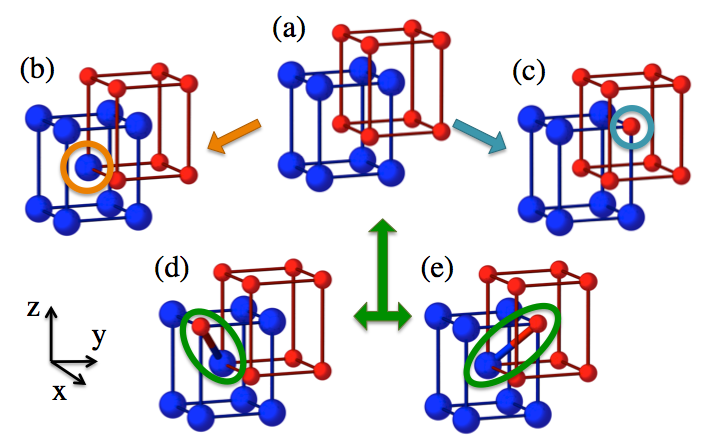}
	\caption{Illustration of considered disorder in FePt. (a) Ideal $L1_0$-ordered FePt. Large blue and small red spheres represent Pt and Fe atoms, respectively. Anti-site defects are introduced by substituting (b) an Fe atom by Pt or (c) vice versa. (d)-(e) Swapping two nearest-neighbor atoms creates a dimer. The bond direction (as indicated by the green ellipse) breaks the tetragonal symmetry of the lattice. In total there are eight differently oriented dimers (two shown).}
	\label{Fig:FePt:defects}
\end{figure}

From a materials perspective, while elemental ferromagnets Fe, Co and Ni give rise to relatively large AHE, they have the disadvantage of weak SOC with corresponding small values of magnetic anisotropy energy \cite{PhysRevLett.87.216405,Handley}. Heavy transition-metals with strong SOC can be successfully doped with magnetic impurities and give rise to large AHE, however, such systems suffer from low Curie temperatures \cite{Crangle:dilute_ferromagnetic_alloys}. The L$1_0$-ordered FePt alloy is by now a classical example of a complex ferromagnet which combines strong SOC and large AHE with strong ferromagnetic ordering. Its crystal structure is depicted in Fig.~\ref{Fig:FePt:defects}a. Remarkably, the strong SOC in combination with uniaxial symmetry of the tetragonal crystal structure leads to a gigantic out-of-plane magneto-crystalline anisotropy energy promising for perpendicular magnetic recording \cite{PhysRevB.71.134411,Farrow:FePt_MAE}, strong anisotropy of the AHE and large anisotropic magnetoresistance~\cite{FePt:Turek}. 

Quite some efforts have been undertaken to analyze the AHE in this material from both  theoretical and experimental sides \cite{AHE_FePt:Hongbin,AnisotropicAHE:Hongbin,AHE_sidejump:Weischenberg,FePt:Czaja}. Seemann \etal \cite{AHE_FePt:Seemann} deduced from a combined experimental and theoretical study that the intrinsic and side-jump contributions to the anomalous Hall conductivity (AHC) are dominant in their samples at elevated temperatures. By extrapolation to zero temperature they were also able to deduce a large magnitude of the skew-scattering Hall angle of 1.10\%. However, experiments by He \etal \cite{FePt:PRL:He} and Chen \etal \cite{FePt-films:APL:Chen,FePt:ChinPhysB:Chen}, report an order of magnitude lower skew-scattering anomalous Hall angles of 0.05\%. Recent \textit{ab initio} calculations, which investigate the effect of long-range order by means of the coherent-potential approximation, find even smaller skew-scattering Hall angles of 0.02\% \cite{FePt:Turek}. In contrast, very large Hall angles of up to 1.5\% have been reported in completely disordered FePt alloys \cite{FePt:ChinPhysB:Chen}. This puzzling situation, as summarized in Table~\ref{Tab:FePt:experiments}, is the starting point of our investigation.

In this Rapid Communication, we show by density functional theory (DFT) calculations that the skew-scattering contribution to the AHE and SHE in L$1_0$ FePt drastically depends on the type of disorder present in real materials. As we show below, simple anti-site defects of Fe and Pt lead to a different sign and magnitude of the AHA, comparable to the large values observed in experiment~\cite{AHE_FePt:Seemann}. In contrast, our values for the AHA in stoichiometric samples with an uncorrelated distribution of defects are considerably lower in magnitude, in line with the previous CPA results \cite{FePt:Turek}. We additionally show, that short-range ordering of defects (that is, a tendency to locate a Pt anti-site defect next to an Fe anti-site defect) has a profound effect on the AHE and SHE, and can change the corresponding Hall angles by a factor of two as compared to the case of uncorrelated disorder.


Our investigations are based on the local spin-density approximation (LSDA) to DFT employing the relativistic full-potential Korringa-Kohn-Rostoker Green function method (rFP-KKR-GF) \cite{Bauer.phd}. In a first step, we obtain the wave-functions  at the Fermi surface of the pristine crystal~\cite{Zimmermann:Fermisurfaces}. Secondly, we self-consistently determine the change of the potential $\Delta V^\mathrm{imp}$ in the presence of a single defect, taking an impurity cluster which contains 19 atoms and takes charge relaxations around the defect into account \cite{Bauer.phd}. As a next step, we calculate the transition rates
\begin{equation}
 P_{\vcc{k} \vcc{k}'} = \frac{2 \pi}{\hbar} N \, c \, \abs{T_{\vcc{k} \vcc{k}'}}^2 \, \delta(E_\vcc{k} - E_{\vcc{k}'}) \komma \label{eq:Pkk}
\end{equation}
where $N$ is the number of atoms in the system, $c$ is the defect concentration, and $T_{\vcc{k} \vcc{k}'}$ is the transition matrix for scattering from a state characterized by the Bloch vector $\vcc{k}$ into a state $\vcc{k}'$ on the Fermi surface (see Ref.~\cite{Long:SHE} for details).
Next, we employ the Boltzmann transport theory to find the vector mean-free path $\boldsymbol{\lambda}(\vcc{k})$ and arrive at equations for the charge-conductivity and spin-conductivity tensors, $\sigma^\mathrm{c}$ and $\sigma^\mathrm{s}$, respectively \cite{Gradhand:SHE:2010,Zimmermann:AHE:2014}.
Note that in Eq.~\eqref{eq:Pkk} the defect concentration enters as a prefactor, which is a good approximation in the dilute limit, where each defect is located far away from other defects and any phase coherence is lost in between two successive scattering events.

\begin{table}[t!]\centering
  \begin{ruledtabular}
    \caption{Summary of literature values for the absolute value of the skew-scattering Hall angle in $L1_0$-ordered FePt alloy. Experimental samples differ in their long-range order, as described by $S$, and film thickness $t$ (in nm).}
	\label{Tab:FePt:experiments}
	\ra{1.2}
    \begin{tabular}{@{}cllll}
      \multicolumn{1}{c}{ $\abs{\alpha^\mathrm{AHE}}$} &
    	\multicolumn{2}{c}{Sample} &
    	\multicolumn{1}{c}{Ref.} \\
 	    \cmidrule[0.4pt](lr{0.125em}){1-4}%
 	1.10\ \% & $S \approx 0.8$,& $t=30$ & Seemann \etal \cite{AHE_FePt:Seemann}  \\
 	0.05\ \% & $S = 0.74$,& $t=10 - 20$ & Chen \etal \cite{FePt-films:APL:Chen}  \\
 	0.05\ \% & $S=0.71$,& $t=20$ & He \etal \cite{FePt:PRL:He}  \\
 	$0.8 - 1.5$\ \% & $S=0$,& $t=30 - 100$ & Chen \etal \cite{FePt:ChinPhysB:Chen} \\
 	0.02\ \% & \multicolumn{2}{c}{TB-LMTO + CPA}  & Kudrnovsky \etal \cite{FePt:Turek}  \\
    \end{tabular}
  \end{ruledtabular}
\end{table}

The $L1_0$ ordered FePt crystal structure can be described by a tetragonal unit cell (lattice constants $a=2.73$\,\AA~and $c/a = 1.39$), where the Pt and Fe atoms are located at $(0,0,0)$ and $(\frac{a}{2},\frac{a}{2},\frac{c}{2})$, respectively [see Fig.~\ref{Fig:FePt:defects}(a)]. The magnetization direction was chosen along the $c$-axis of the crystal~\footnote{Note that the sign of $\sigma_{xy}$ depends on the direction of the magnetization direction. Here, the magnetization direction (spin moment) is chosen along the $+z$ ($-z$)-direction of the coordinate system.}, as this is the easy-axis determined by the magneto-crystalline anisotropy energy \cite{Staunton.2004.FePtL10}. The calculation of the total spin moment provides a value of $3.21 \, \mu_\mathrm{B}$ per unit cell, of which the Fe and Pt atoms contribute $2.88 \, \mu_\mathrm{B}$ and $0.33 \, \mu_\mathrm{B}$, respectively.


\begin{table*}[htb!]\centering
 \begin{ruledtabular}
 	\caption{Disorder-induced transport properties of the $L1_0$-ordered FePt alloy. Conductivities are given in units of $10^6~\mathrm{S/m} = (\mu \Omega \, \mathrm{m})^{-1}$ at an impurity concentration of $1\,\mathrm{at.}\%$. Hall angles are given in percent.}
 	\label{Tab:FePt:results}
 	\ra{1.2}
	\begin{tabular}{@{}ld{1}d{1}d{3}d{3}d{2}d{2}@{}}
		&
		\multicolumn{1}{c}{$\sigma^\mathrm{c}_{xx}$} &
		\multicolumn{1}{c}{$\sigma^\mathrm{c}_{zz}$} &
		\multicolumn{1}{c}{$\sigma^\mathrm{c}_{xy}$} &
		\multicolumn{1}{c}{$\sigma^\mathrm{s}_{xy}$} &
		\multicolumn{1}{c}{$\alpha^\mathrm{AHE}$} &
		\multicolumn{1}{c}{$\alpha^\mathrm{SHE}$} \\
		\cmidrule[0.4pt](lr{0.125em}){2-3}%
		\cmidrule[0.4pt](lr{0.125em}){4-5}%
		\cmidrule[0.4pt](l{0.125em}){6-7}%
		Pt impurity ($\mathrm{Fe}_{0.99}\mathrm{Pt}_{1.01}$)  &  64.5 &  33.8 & 0.175 & 0.141 & 0.27 & 0.21 \\
		Fe impurity ($\mathrm{Fe}_{1.01}\mathrm{Pt}_{0.99}$)  & 131.7 & 111.8 &  -1.15  &  -0.883 &  -0.88 &  -0.67 \\
		uncorr. mixture ($S = 0.99$) &  70.7 &  46.3 &  -0.064 &  -0.093 &  -0.09 &  -0.13 \\
		dimer (SRO, 1\% defect atoms)                         &  76.9 &  51.4 &  -0.118 &  -0.057 &  -0.15 &  -0.07 \\

	\end{tabular}
\end{ruledtabular}
\end{table*}

First we discuss the anti-site defects, which is a substitution of Fe atoms by Pt (further for simplicity termed Pt \emph{impurity}) and vice versa, to simulate a weak deviation from the stoichiometric ratio [\ie $\mathrm{Fe}_{1-c}\mathrm{Pt}_{1+c}$, see Figs.~1(b) and 1(c)]. We work in the dilute limit, \ie small impurity concentrations $c$, in which $\alpha^\mathrm{AHE}$ and $\alpha^\mathrm{SHE}$ are actually independent of $c$. However, absolute values for the conductivity scale inversely proportional to $c$, and we give values corresponding to $c=1\%$.

\begin{figure}[t!]
	\includegraphics[width=0.42\textwidth]{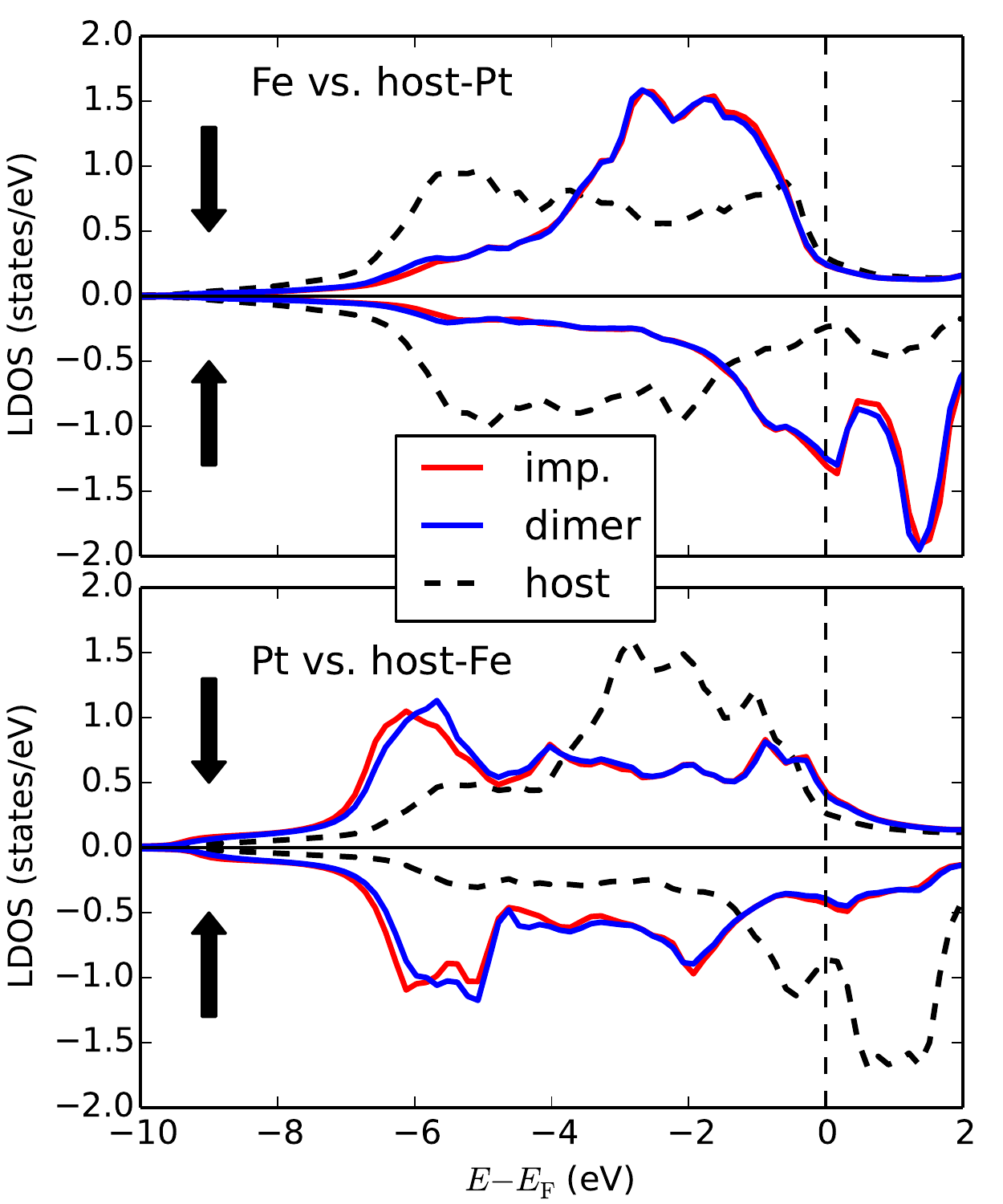}
	\caption{Spin-resolved local density of states of an Fe impurity and the Fe-atom in a dimer as compared to the substituted host Pt-atom (upper panel) and vice versa (lower panel). Arrows pointing downwards and upwards correspond to majoriy and minority spins, respectively \cite{Note1}.}
	\label{Fig:dos}
\end{figure}

Our main results are shown in Table~\ref{Tab:FePt:results}. The non-stoichiometric alloys show very different transport properties. Their longitudinal conductivities ($\sigma^\mathrm{c}_{xx}$ and $\sigma^\mathrm{c}_{zz}$) differ by a factor of $2-3$, being larger for Fe impurities than for Pt impurities. This shows that Pt impurities are more effective scattering centers to the incoming Bloch electrons, or in other words, that Fe impurities are more ``transparent'' to the propagating electrons. This fact is consistent with the local density of states (LDOS) of an Fe impurity compared to a host-Pt atom which it substitutes [see Fig.~\ref{Fig:dos}(a)]: Incoming majority electrons at the Fermi level see an Fe-impurity LDOS which is nearly indistinguishable from the one of a host-Pt atom and scatter very weakly [in numbers, $n^\downarrow_\mathrm{F}=0.25$ ($0.29$) states/eV for an Fe impurity (host Pt)]. Indeed, from our full calculations we find that majority electrons constitute most of the conductivity, resulting in a 50\% spin-polarization of the diagonal conductivity. Such a strong difference in the influence of disorder on transport properties of majority versus minority electrons is very common in magnetic materials, and is responsible for \eg an anomalous concentration dependence of the resistivity of Fe upon alloying with Cr or V \cite{Ebert_transport_in_ferromagnets:PRB2009}. On the contrary, the LDOS of a Pt impurity deviates around the Fermi energy considerably from the substituted host-Fe atom in both spin channels [\cf lower panel of Fig.~\ref{Fig:dos}, $n^\downarrow_\mathrm{F}=0.43$ (0.26) states/eV for a Pt impurity (host Fe)], and electrons scatter much stronger, which is also in line with a reduction of spin-polarization to 36\%. As far as the transverse transport properties are concerned, the AHCs can be quite large, leading to a sizable anomalous Hall angle of $\alpha^\mathrm{AHE} = 0.27\%$ for Pt impurities (see Fig.~\ref{Fig:FePt:defects}b). Switching to Fe impurities, $\alpha^\mathrm{AHE}$ grows in magnitude and even changes its sign to constitute a large value of $\alpha^\mathrm{AHE}=-0.88\%$. This shows the drastic influence of different types of disorder on the skew-scattering contribution in FePt.
The spin-Hall conductivity (SHC) to good extent follows the AHC values (see Table~\ref{Tab:FePt:results}). This implies that the transverse current is strongly spin-polarized, with a spin-polarization of 80\% (77\%) for Pt (Fe) impurities.

Next, we simulate weak uncorrelated disorder, keeping the stoichiometric ratio but replacing in equal amounts some Fe atoms by Pt and vice versa. Assuming that electrons scatter independently off the two types of impurities, we can use the previously calculated transition rates and average them as~\cite{Mertig:Matthiessen_rule}
\begin{equation}
  P^\mathrm{avg}_{\vcc{k}\vcc{k}'} = P^\mathrm{Fe}_{\vcc{k}\vcc{k}'} + P^\mathrm{Pt}_{\vcc{k}\vcc{k}'} \komma \label{Eq:Pkkaverage}
\end{equation}
where the concentrations for the two types of impurities are determined by the long-range order parameter $S$ according to $c_\mathrm{Fe} = c_\mathrm{Pt} = (1-S)/2$. 
For this uncorrelated mixture of defects, the values of the diagonal conductivities are in between the ones of the previously discussed non-stoichiometric crystals, being much closer to the Pt-impurity case. Both, the transverse charge and spin conductivities are significantly reduced as compared to the previous cases (see Tab.~\ref{Tab:FePt:results}). A heuristic argument for this reduction is a partial compensation of the opposite in sign skew-scattering off Fe and Pt impurities. An anomalous Hall angle of approx.~$-0.11\%$ deduced from very recent TB-LMTO-CPA calculations \cite{Hyodo:FePt-LMTO-CPA} is in very good agreement to our value of $-0.09\%$ (see Table~\ref{Tab:FePt:results}).

Let us now see whether we can arrive at this result by simpler means. First, it is seemingly plausible to think of electrons of opposite spin (up, $\uparrow$, and down, $\downarrow$) as distinct entities which do not interact with each other and which separately contribute to the charge and spin conductivity: $\sigma^\mathrm{c} = \sigma^{\uparrow} + \sigma^{\downarrow}$ and $\sigma^\mathrm{s}=\sigma^{\uparrow} - \sigma^{\downarrow}$ [In fact, we used this picture above in our interpretation of diagonal conductivities in terms of the LDOS]. The Matthiessen rule states that the resistivities
can be simply added if the two scattering sources are independent of each other,\,i.e.
\begin{equation}
 \left( \sigma^\uparrow \right)^{-1} = \frac{1}{2} \left[ \left( \sigma^\uparrow_\mathrm{Fe} \right)^{-1} + \left( \sigma^\uparrow_\mathrm{Pt} \right)^{-1} \right] \komma \label{Eq:average_tensor}
\end{equation}
and similarly for $\sigma^\downarrow$. This procedure yields values for the elements of the averaged charge-conductivity tensor that are about 20\% too high in magnitude compared to the full calculation. In contrast, the transverse spin-conductivity comes out by a factor of 8 too small. This discrepancy originates in the fact that due to the strong spin-orbit coupling in FePt the electronic wavefunctions are 
strongly spin-mixed and the two-current ansatz evidently fails in this case. 

An alternative approach would be to regard the charge- and spin-currents independently, and to perform the averaging in analogy to Eq.~\eqref{Eq:average_tensor} directly on the level of the charge and spin-conductivity tensors. This again gives a reasonable estimate for the elements of the charge conductivity tensor, but as far as the transverse spin conductivity is concerned, not even the sign of it can be reproduced correctly ($\sigma_{xy}^\mathrm{s}=400~\mathrm{S}/\mathrm{cm}$ as compared to $-930~\mathrm{S}/\mathrm{cm}$ for the full calculation). In conclusion, the Matthiessen rule approximations work quite well for charge transport, but greatly fail for spin-transport properties of L$1_0$ FePt alloy.

Eq.~\eqref{Eq:Pkkaverage} entails the approximations that (i) the wavefunction phase is lost due to random positions of the impurities and (ii) the concentration is small enough that multiple-scattering effects between impurities can be neglected. However, in case of correlated impurity positions
these approximations are not valid anymore. In order to estimate the impact of such effects on transport properties, we investigate the extremal case of two anti-site defects being nearest neighbors,\,i.e.\,when nearest-neighbor Fe and Pt atoms swap their positions and form a dimer [Fig.~\ref{Fig:FePt:defects}(d)-(e)]. This class of defects simulates an ultimate case of short-range ordering (SRO) of defects. Generally, there are eight possible orientations of the dimer bond, and in a realistic situation they would appear with equal probability randomly distributed over the crystal.
We emphasize that we perform a full calculation for each dimer orientation, \ie we swap the two atoms in the impurity cluster, calculate the self-consistent impurity potentials, and finally obtain the transition rates directly from Eq.~\eqref{eq:Pkk}. Next, we average over the dimer orientations on the level of the transition rates in analogy to Eq.~\eqref{Eq:Pkkaverage}, which neglects dimer-dimer interference effects. We choose the concentrations such that in total 1\% of the crystal sites are defects.

Comparing first the local density of states (LDOS) of an Fe-atom in the dimer to a simple anti-site Fe impurity (see upper panel of Fig.~\ref{Fig:dos}), we remark that the two LDOS are practically identical. The same is true for the Pt atom in the dimer compared to a Pt impurity (see lower panel), with minor modifications of the occupied states around 6~eV below the Fermi level. This similarity could suggest very similar transport properties between the uncorrelated mixture and the SRO case. Indeed, the full calculation reveals that SRO increases the diagonal conductivity by only 10\% as compared to long-range disorder (see Table.~\ref{Tab:FePt:results}). This is qualitatively in line with Ref.~\cite{Tulip_shortrangeorder}, where a moderate decrease in the longitudinal resistivity upon inclusion of SRO in CuZn-alloys was predicted from calculations based on the non-local coherent potential approximation.

On the contrary, SRO has a profound impact on the transverse transport properties (see Table~\ref{Tab:FePt:results}). Interestingly, $\sigma^\mathrm{c}_{xy}$ is increased by a factor of roughly $2$, whereas $\sigma^\mathrm{s}_{xy}$ is reduced by a factor of $1.6$, with similar trends for the anomalous and spin-Hall angles. Our results show that transverse transport properties depend on the fine details of scattering at the Fermi surface, and full \emph{ab-initio} calculations are required to describe complex disorder reliably.



To summarize, we have shown that the skew-scattering anomalous Hall and spin-Hall angles of L$1_0$-ordered FePt drastically depend on the disorder type. Remarkably, the sign of the AHE is changed when the composition of the alloy slightly deviates from the stoichiometric ratio towards the Fe-rich side as compared to the Pt-rich side. Short-range ordering of defects has a profound effect on the Hall angles and can change them by a factor of $2$ as compared to the case of dilute uncorrelated disorder. This might explain the vast range of anomalous Hall angles measured in experiments on different samples of this alloy, which undergo different preparation procedures and differ in their crystallographic quality. The detailed microscopic understanding of skew-scattering in such alloys paves the way towards an educated ability of engineering the desired Hall transport properties of transition metals.


\bigskip

We acknowledge discussions with I.~Turek, as well as funding under DFG Project No.~SPP~1538 ``Spin-caloric Transport'', as well as computing time at the J\"ulich Supercomputing Centre and JARA-HPC of RWTH Aachen University.

\bibliography{ref}

\end{document}